\documentclass[nofootinbib,prd,showpacs]{revtex4}
\usepackage{amsmath}
\usepackage{amsfonts}
\usepackage{amssymb}
\usepackage{graphicx}

\begin{document}

\title{Ideal--Modified Bosonic Gas Trapped in an Arbitrary 3--dim Power--Law Potential}
\author{E. Castellanos}
\email{elias@zarm.uni-bremen.de}\affiliation{ZARM, Universit\"at
Bremen, Am Fallturm, 28359 Bremen, Germany}

\author{C. L\"ammerzahl}
\email{laemmerzahl@zarm.uni-bremen.de} \affiliation{ZARM,
Universit\"at Bremen, Am Fallturm, 28359 Bremen, Germany}

\date{\today}

\begin{abstract}
We analyze the effects caused by an anomalous single--particle
dispersion relation suggested in several quantum--gravity models,
upon the thermodynamics of a Bose--Einstein condensate trapped in a generic
3-dimensional power--law potential. We prove that the shift in the
condensation temperature, caused by a deformed dispersion relation,
described as a non--trivial function of the number of particles and the shape associated to
the corresponding trap, could provide bounds for the parameters
associated to such deformation. Additionally, we calculate the fluctuations in the number of particles as a criterium of thermodynamic stability for these systems. We show that the apparent
instability caused by the \emph{anomalous} fluctuations in the
thermodynamic limit can be suppressed considering the lowest
energy associated to the system in question.
\end{abstract}

\maketitle

\section{Introduction}

The search for small manifestations of quantum gravity or quantum
gravity phenomenology in our low energy world is a very
controversial topic in modern physics. In some schemes, the
possibility that the space-time could be quantized, can be
characterized, from a phenomenological point of view, as a
modification in the dispersion relation of microscopic particles
\cite{Giovanni1,Kostelecky,Claus} (and references therein). A modified dispersion relation
emerges as an adequate tool in the search for phenomenological
consequences caused by this type of quantum gravity models.
Nevertheless, the principal difficulty in the search of
manifestations of quantum gravity is the smallness in the predicted
effects \cite{Kostelecky,amelino1}. If this kind of deformations are
characterized by some Planck scale, then the quantum gravity effects
become very small \cite{Giovanni1,Claus}. In the non--relativistic
limit, the dispersion relation can be expressed as follows
\cite{Claus,Claus1}
\begin{equation}
E \simeq
m+\frac{p^{2}}{2m}+\frac{1}{2M_{p}}\Bigl(\xi_{1}mp+\xi_{2}p^{2}+\xi_{3}\frac{p^{2}}{m}\Bigr).
\label{ddr}
\end{equation}
Equation (\ref{ddr}) is expressed in units where the speed of light
$c=1$, being $M_{p}$ ($\simeq 1.2\times 10^{28}eV$) the Planck mass.
The three parameters $\xi_{1}$, $\xi_{2}$, and $\xi_{3}$, are model
dependent \cite{Giovanni1,Claus}, and should take positive or
negative values close to $1$. There are some evidence within the
formalism of Loop quantum gravity \cite{Claus,Claus1,5,12} that
indicates a non--zero values for the three parameters, $\xi_{1},\,
\xi_{2},\, \xi_{3}$, and particulary \cite{5,13} that produces a
linear--momentum term in the non--relativistic limit. Unfortunately,
as is usual in quantum gravity phenomenology, the possible bounds
associated with the deformation parameters, open a wide range of
possible magnitudes, which is translated to a significant challenge.
On the other hand, the Bose--Einstein condensation phenomenon, from
the theoretical and experimental point of view, has produced an
enormous amount of interesting publications associated to this topic
\cite{Dalfovo,bagnato,grossmann,Giorgini,Haugerud,Li,zijun,Salasnich,Zobay,Jaouadi,ketterle,Politzer,RP,grossmann1,grossmann2,yukalov,yukalov1,Vitaly}
(and references therein). Among the issues addressed we may find its
possible use as tools in the search of quantum--gravity
manifestations, for instance, in the context of Lorentz violation or
to provide phenomenological constrains on Planck--scale physics
\cite{Claus,Colladay,Donald,Camacho,CastellanosCamacho,CastellanosCamacho1,CastellanosClaus,r1,r2}.
In this sense, it is rather exciting, to look at the effects over
the thermodynamic properties associated to the Bose--Einstein
condensates (which itself, is a manifestation of quantum effects at
macroscopic level), caused by the quantum structure of the
space--time. For this purpose, we define the next modified
Hamiltonian
\begin{equation}
H=\frac{p^{2}}{2m}+\alpha p+U(\vec{r}) \label{defH}
\end{equation}
where $p$ is the momentum, $m$ is the mass of the particle, and the
term $\alpha p$, with $\alpha=\xi_{1}\frac{m}{2M_{p}}c$ in ordinary
units, is the leading order modification in expression (\ref{ddr}),
being $c$ the speed of light.

The potential term
\begin{equation}
\label{potgen}
 U(\vec{r})=\sum_{i=1} ^ {d} A_{i}\Bigg|\frac{r_{i}}{a_{i}}\Bigg|^{s_{i}},
\end{equation}
is the generic 3--dimensional power--law potential, where $ A_{i}$
and $a_{i}$ are energy and length scales associated to the trap
\cite{Jaouadi}. Additionally, $r_{i}$ are the $d$ radial coordinates
in the $n_{i}$--dimensional subspace of the 3--dimensional space.
The sub--dimensions $n_{i}$ satisfy the following expression in
three spatial dimensions
\begin{equation}
\sum_{i=1} ^ {d} n_{i}=3. \label{subdim}
\end{equation}
If in equation (\ref{subdim}) $d=3$, $n_{1}=n_{2}=n_{3}=1$, then the
potential becomes in the Cartesian trap. If $d=2$, $n_{1}=2$ and
$n_{2}=1$, then we obtain the cylindrical trap. If $d=1$, $n_{1}=3$,
then we have the spherical trap. If $s_{i}\rightarrow\infty$, we
have a free gas in a box. In this sense, the potential included in
the Hamiltonian (\ref{defH}) is quite general. Different
combinations of these parameters give different classes of
potentials, according to (\ref{potgen}). It is noteworthy to mention
that the use of these generic potentials, opens the possibility to
adiabatically cool the system in a reversible way, by changing the
shape of the trap \cite{Dalfovo}.
The analysis of a Bose--Einstein condensate in the ideal case,
weakly interacting, and with a finite number of particles, trapped
in different potentials shows that the main properties associated to
the condensate, and in particular the critical temperature, depends
strongly on the trapping potential in question
\cite{bagnato,grossmann,Giorgini,ketterle,Haugerud,Li,zijun,Salasnich,Zobay,Jaouadi}.
The characteristics of the potential (in particular, the parameter
that defines the shape of the potential) has a strong impact on the
dependence of the critical temperature with the number of particles
(or the associated density).
The main objective of this work, is analyze the thermodynamics of a
Bose--Einstein condensate with an anomalous single--particle
dispersion relation, in order to show that this system, trapped in
a quite general potential, can be used to establish some meaningful
bounds for the parameter $\xi_{1}$.

\section{Condensation Temperature}
The first part of the present work addresses the issue of the
effects of a deformed dispersion relation over the condensation
temperature for an ideal bosonic gas, trapped  in the generic
potential given by expression (\ref{potgen}) within the semiclassical
approximation \cite{Dalfovo,Pethick}.
The semiclassical energy associated with the modified Hamiltonian
(\ref{defH}) is given by
\begin{equation}
\label{HF} \epsilon(\vec{r}, \vec{p})=\frac{p^{2}}{2m}+ \alpha
p+U(\vec{r}).
\end{equation}
In the semiclassical approximation, the single--particle
phase--space distribution may be written as \cite{Dalfovo,Pethick}
\begin{equation}
\label{SPSD}
n(\vec{r},\vec{p})=\frac{1}{e^{\beta(\epsilon(\vec{r},\vec{p})-\mu)}-1},
\end{equation}
where $\beta=1/\kappa T$, $\kappa$ is the Boltzmann constant, and
$T$ the temperature. Additionally, $\mu$ is the chemical potential.
The number of particles in the 3--dimensional space obeys the
normalization condition \cite{Dalfovo,Pethick},
\begin{equation}
 N=\frac{1}{(2 \pi \hbar)^{3} }\int d^{3}\vec{r}\hspace{0.1cm} d^{3} \vec{p}\hspace{0.1cm} n(\vec{r},\vec{p}),
\label{NC}
\end{equation}
where
\begin{equation}
\label{n1} n(\vec{r})=\int  d^{3} \vec{p} \hspace{0.1cm}
n(\vec{r},\vec{p}),
\end{equation}
is the spatial density \cite{Dalfovo,Pethick}. Using expression
(\ref{HF}), and integrating expression (\ref{SPSD}) over the
momentum space, with the help of (\ref{n1}), leads us to the spatial
distribution associated with the modified semi--classical spectrum
(\ref{HF})
\begin{eqnarray}
\label{MSD}
 n(\vec{r})&=&\lambda^{-3} g_{3/2}\Bigl(e^{\beta(\mu_{eff}-U(\vec{r}))}\Bigr)-
\alpha \lambda^{-2}\Bigl(\frac{m}{\pi \hbar}\Bigr)
g_{1}\Bigl(e^{\beta(\mu_{eff}-U(\vec{r}))}\Bigr)
+\alpha^{2}\lambda^{-1}\Bigl(\frac{m^{2}}{2\pi
\hbar^{2}}\Bigr)g_{1/2}\Bigl(e^{\beta(\mu_{eff}-U(\vec{r}))}\Bigr),
\end{eqnarray}
where $\lambda=\Bigl(\frac{2 \pi \hbar^{2}}{m \kappa
T}\Bigr)^{1/2}$, is the de Broglie thermal wavelength,
$\mu_{eff}=\mu+m\alpha^{2}/2$ is an effective chemical potential,
and $g_{\nu}(z)$ is the so--called Bose--Einstein function defined
by \cite{Phatria}
\begin{equation}
g_{\nu}(z)=\frac{1}{\Gamma(\nu)}\int_{0}^{\infty}\frac{x^{\nu-
1}dx}{z^{-1}e^{x}-1}. \label{BEF}
\end{equation}
The Bose--Einstein function (\ref{BEF}) diverges for $z=1$ when $\nu
\leq 1$ \cite{Phatria}. The behavior of the Bose--Einstein functions
and its relation with the value of the chemical potential at the
critical temperature is usually used as a criterium of condensation
\cite{yukalov1,Phatria}.
If we set $\alpha=0$ in equation (\ref{MSD}) we recover the usual
result for the spatial density in the semiclassical approximation
\cite{Dalfovo,Pethick}.
Integrating the normalization condition (\ref{NC}), using expression
(\ref{MSD}) with the corresponding potential (\ref{potgen})
allows to obtain an expression for the number of particles as a
function of the chemical potential $\mu$, the temperature $T$, and
the deformation parameter $\alpha$. Additionally, accepting that
$m\alpha^{2}/2 << \kappa T$, allows to obtain an expression for the number
of particles $N$ at first order in $\alpha$, by using the properties
of the Bose--Einstein functions \cite{Phatria}, with the result
\begin{eqnarray}
\label{NPINT1}
N=N_{0}&+&C\Pi_{l=1}^{d}A_{l}^{-\frac{n_{l}}{s_{l}}}a_{l}^{n_{l}}\Gamma
\Bigl(\frac{n_{l}}{s_{l}}+1\Bigr)\Bigg[\Bigl(\frac{m}{2\pi
\hbar^{2}}\Bigr)^{3/2} g_{\gamma}(z)(\kappa T)^{\gamma}-\alpha
\Bigl(\frac{m^2}{2\pi^{2} \hbar^{3}}\Bigr)g_{\gamma-1/2}(z)(\kappa
T)^{\gamma-1/2}\Bigg].
\end{eqnarray}
where
\begin{equation}
\label{parametergamma}
\gamma=\frac{3}{2}+\sum_{l=1}^{d}\frac{n_{l}}{s_{l}},
\end{equation}
is the parameter that defines the shape of the potential
(\ref{potgen}). In the case of a harmonic oscillator in three
dimensions $\gamma=3$. Additionally, $N_{0}$ are the particles in the ground
state, $z=e^{\beta\mu}$ is the so--called fugacity, and $\Gamma(y)$ is the Gamma function. The constant $C$ depends on the potential in question and in the case of Cartesian
traps, and in consequence, for a three dimensional harmonic
oscillator potential $C=8$. Setting $\alpha=0$ in (\ref{NPINT1}) we
recover the result given in \cite{Jaouadi}.
At the critical temperature in the thermodynamic limit the fugacity $z\approx1$ (or equivalently $\mu\approx 0$) and $N_{0}\approx 0$, which implies that the Bose--Einstein functions are given
by the corresponding Riemann Zeta functions $\zeta(x)$
\cite{Phatria}, then from expression (\ref{NPINT1}) we obtain that the number of particles at
the condensation temperature is given by
\begin{eqnarray}
\label{temcrit}
N&=&C\Pi_{l=1}^{d}A_{l}^{-\frac{n_{l}}{s_{l}}}a_{l}^{n_{l}}\Gamma
\Bigl(\frac{n_{l}}{s_{l}}+1\Bigr)\Bigg[\Bigl(\frac{m}{2\pi
\hbar^{2}}\Bigr)^{3/2} \zeta (\gamma)(\kappa T_{c})^{\gamma} -\alpha
\Bigl(\frac{m^2}{2\pi^{2} \hbar^{3}}\Bigr)\zeta(\gamma-1/2)(\kappa
T_{c})^{\gamma-\frac{1}{2}}\Bigg],
\end{eqnarray}
where $T_{c}$ is the condensation temperature. If we set $\alpha=0$ in
(\ref{temcrit}), we recover the usual expression for the condensation
temperature $T_{0}$ \cite{Jaouadi}
\begin{equation}
\label{criter}
T_{0}=\Bigg[\frac{N\Pi_{l=1}^{d}A_{l}^{\frac{n_{l}}{s_{l}}}a_{l}^{-n_{l}}}{C\Pi_{l=1}^{d}\Gamma
\Bigl(\frac{n_{l}}{s_{l}}+1\Bigr)\zeta(\gamma)}\Bigl(\frac{2 \pi
\hbar^{2}}{m}\Bigr)^{3/2}\Bigg]^{1/\gamma}\frac{1}{\kappa}.
\end{equation}
Let us define
\begin{equation}
\label{volumen}
V_{char}=\frac{\Pi_{l=1}^{d}A_{l}^{\frac{n_{l}}{s_{l}}}a_{l}^{-n_{l}}}{C\Pi_{l=1}^{d}\Gamma
\Bigl(\frac{n_{l}}{s_{l}}+1\Bigr)},
\end{equation}
as the characteristic volume associated with the system. From
expression (\ref{volumen}) we notice that if
$s_{i}\rightarrow\infty$ then, $ V_{char}$ becomes the volume
associated with a free gas in a box (in fact the inverse of the
volume with this definition). In this sense, $V_{char}$ can be
interpreted as the available volume occupied by the gas
\cite{zijun,yukalov1}. At this point, it is noteworthy to mention that
the most general definition of thermodynamic limit can be expressed
as follows
\begin{equation}
\label{TLi}
 N\rightarrow \infty,\hspace{0.5cm}V_{char}\rightarrow0,
\end{equation}
keeping the product $NV_{char}\rightarrow const$, and is valid for
all power law potentials in any spatial dimensionality
\cite{yukalov1}. With the criterium given above, the critical
temperature in the thermodynamic limit is well defined
\cite{Dalfovo}.
From equations (\ref{temcrit}) and (\ref{criter}), we obtain the
shift in the critical temperature for our deformed bosonic gas
\begin{equation}
 \label{shift1}
\Bigl(\kappa T_{c}\Bigr)^{\gamma}=\Bigl(\kappa T_{0}\Bigr)^{\gamma}
+\alpha \Bigl(\frac{2m}{\pi}\Bigr)^{1/2}\frac{\zeta(\gamma-1/2)}
{\zeta(\gamma)}\Bigl(\kappa T_{c}\Bigr)^{\gamma-1/2}.
\end{equation}
From expression (\ref{shift1}), we notice that $T_{c}$ increases for
$\alpha>0$, compared with the corresponding critical temperature of
the usual Bosonic gas when $\alpha=0$. The opposite case,
$\alpha<0$, corresponds to a decrease in the critical temperature,
compared with the usual case. Clearly, if $\alpha=0$, then
$T_{c}=T_{0}$. The effect of the external potential is to
concentrate particles in the center of the trap and a positive
$\alpha$ increases the critical temperature, which means that, the
effect of $\alpha>0$ reinforce the main effect of the external
potential. When $\alpha<0$ we have the opposite behavior, a negative
$\alpha$ tends to weaken the effect of the external potential. We
can also express the shift in the critical temperature as a function
of the number of particles $N$,
\begin{equation}
\label{CTNP} \frac{T_{c}-T_{0}}{T_{0}}\equiv\frac{\Delta
T_{c}}{T_{0}}\simeq\alpha \Omega N^{-1/2\gamma},
\end{equation}
where
\begin{equation}\label{omi}
\Omega=\Bigg(\frac{2m}{\pi}\Bigg)^{1/2}\frac{\zeta(\gamma-1/2)}{\gamma\zeta(\gamma)}
\Bigg(\frac{V_{char}(2\hbar^{2})^{3/2}}{\zeta(\gamma)}\Bigg)^{-1/2\gamma}.
\end{equation}
From (\ref{CTNP}) and (\ref{omi}), we notice that correction in the
condensation temperature depends strongly on the functional form between
the number of particles and the parameters associated to the
potential in question. In typical experiments the number of
particles vary from a few thousand to several millions, and
frequencies $ \bar{\omega}/2\pi $ from tens to hundreds of Hertz,
together with values of $10^{13}$ to $10^{15}$ atoms per $cm^{-3}$
\cite{Dalfovo}. Additionally, the current high precision experiments
in the case of $^{39}_{19} K$, the shift in the condensation temperature
respect to the ideal result, caused by the interactions among the
constituents of the gas is about $5\times 10^{-2}$ with a $1\%$  of
error \cite{RP}. These facts, allows us to compare the results given
here, in order to obtain bounds for the deformation parameter
$\xi_{1}$. For the sake of simplicity, let us analyze the case of
spherical traps, in such a case the corresponding potential is given
by $U(r)=A_{1}(\frac{r}{a_{1}})^{s_{1}}$, where $A_{1}=\hbar
\omega_{0}/2$ and $a_{1}=\sqrt{\hbar/m \omega_{0}}$ [see
expression (\ref{volumen})].
In this case, the shift in the critical temperature is given by
\begin{equation}
\label{esferico} \frac{\Delta T_{c}}{T_{0}}\simeq\alpha
\Omega_{s_{1}} N^{-s_{1}/3(s_{1}+2)}.
\end{equation}
For different values of $s_{1}$ we obtain, $\frac{\Delta
T_{c}}{T_{0}}\sim\alpha N^{-1/9}$ , for $s_{1}=1$, which corresponds
to a linear trap. For $s_{1}=2$, $\frac{\Delta
T_{c}}{T_{0}}\sim\alpha N^{-1/6}$, which is an isotropic harmonic
oscillator. For  $s_{1}=3$,  $\frac{\Delta T_{c}}{T_{0}}\sim\alpha
N^{-1/5}$. For  $s_{1}=6$,  $\frac{\Delta T_{c}}{T_{0}}\sim\alpha
N^{-1/4}$, and so on. We noticed immediately that if
$s_{1}\rightarrow \infty$, after some algebraic manipulation then,
we are able to obtain the limiting case of a bosonic gas trapped in
a box, in such a case,
\begin{equation}
\frac{\Delta
T_{c}}{T_{0}}\simeq\alpha\frac{2m(V\zeta(3))^{1/3}}{3\hbar}N^{-1/3}.
\end{equation}
For these values of the parameter $s_{1}$, we obtain a bound for the
deformation parameter $|\xi_{1}|\lesssim10^{6}$ for the linear trap
$s_{1}=1$ (with frequencies $\omega_{0}\sim 15\, Hz$ and $N\sim
10^{6}$), to $|\xi_{1}|\lesssim10^{2}$ corresponding to a free gas
in a box $s_{1}\rightarrow \infty$, with densities about
$10^{13}-10^{15}$, for $^{39}_{19} K$, with a mass $15 \times 10^{-26}\,Kg$.
In fact, these bounds could be improved in a system containing a small number of
particles and/or a massive bosons and/or lower frequencies but,
where the thermodynamic limit is still valid, trapped in potentials
where the parameter $s_{1}$ is sufficiently large. Additionally, we are
able to improve the bound associated to the deformation parameter
$\xi_{1}$ by the use of different classes of potentials and it is
straightforward to generalize this result to a more general
potentials by using (\ref{CTNP}).

Because the relevance of harmonic traps in the experiments, let us
consider the case in which our condensate is trapped in an
anisotropic three-dimensional harmonic--oscillator potential. For
this trap, the shape parameter is given by $\gamma=3$ with
$A_{i}=\hbar\omega_{i}/2$ and $a_{i}=\sqrt{\hbar/m\omega_{i}}$ [see
expression (\ref{volumen})]. Using the definition
$\bar{\omega}=(\omega_{1}\omega_{2}\omega_{3})^{1/3}$ expression
(\ref{CTNP}) becomes
\begin{equation}
\label{corearm} \frac{\Delta T_{c}}{T_{0}}\simeq
\alpha\frac{\zeta(5/2)}{3\zeta(3)^{5/6}}
\Bigg(\frac{8m}{\hbar\bar{\omega} \pi }\Bigg)^{1/2}N^{-1/6}.
\end{equation}
In this case, with say, $\omega_{1}\sim10$Hz, $\omega_{2}\sim10$Hz,
$\omega_{3}\sim20$Hz, and $\alpha=\xi_{1}\frac{m}{2M_{p}}c$, $c$ is
the speed of light and $M_{p}$ is the Planck's mass, with say,
$N\sim10^{18}$, $N\sim10^{9}$, and $N\sim10^{6}$, we obtain from
(\ref{corearm}) that the shift in the critical temperature is
approximately given by $\xi_{1}\,10^{-8}$, $\xi_{1}\,10^{-7}$,
and $\xi_{1}\,10^{-6}$.  In this situation, by using the experimental
data given above, we obtain for the deformation parameter $\xi_{1}$,
$|\xi_{1}| \lesssim 10^{6}$, $|\xi_{1}|\lesssim10^{5}$, and
$|\xi_{1}|\lesssim10^{4}$, for  $N\sim10^{18}$, $N\sim10^{9}$, and
$N\sim10^{6}$ respectively. These results show that we are able to
improve the bounds for the deformation parameter by varying the
number of particles and/or the frequencies associated to the trap.
These cases, illustrate how a Bose--Einstein condensate, could be
used, in principle, to provide bounds for the deformation parameter
$\xi_{1}$ suggested in several quantum--gravity models. In
references \cite{Claus,Claus1} it is suggested the use of
ultra-precise cold-atom-recoil experiments to constrain the form of
the energy-momentum dispersion relation in the non--relativistic
limit or at low energies. In \cite{Claus,Claus1}, the bound
associated to $\xi_{1}$ is at less, four orders of magnitude smaller
than the bonds obtained in the case of an anisotropic three-dimensional
harmonic--oscillator potential. Nevertheless, the use of
Bose--Einstein condensates in this context is noticeable and opens
the possibility to improve these bounds, in systems containing a
finite number of particles, in the non--interacting and interacting
cases.

The study of fluctuations in the number of particles within the
Bose--Einstein condensation phenomenon is important for several reasons
\cite{Politzer,yukalov,yukalov1,yukalov3,Vitaly} (and references therein). For instance, the
fluctuations in the number of particles are directly related with
the equivalence or non--equivalence of the statistical ensembles
\cite{Politzer,Vitaly} which is a non--trivial and deep topic. On
the other hand, the fluctuations in the number of particles are
directly related with the thermodynamical stability of the system in
question \cite{yukalov,yukalov1,yukalov3}. The fluctuations in a
Bose--Einstein condensate depend strongly on the single--particle
energy spectrum \cite{Vitaly} and consequently the appearance of the
parameter $\alpha$ in expression (\ref{shift1}) or (\ref{CTNP}),
could affect the thermodynamic stability of the system. One way of
quantifying the stability of the system is through the fluctuations
in the number of particles, which are directly related to the
so--called isothermal compressibility. As a consequence, a stable
system requires the fluctuations in the number of particles to be
necessarily \emph{normal} in the thermodynamic limit, i.e.,
proportional to the number of particles
\cite{yukalov,yukalov1,yukalov3}.
The number of particle fluctuations are characterized by the
dispersion \cite{yukalov,yukalov1,Phatria}
\begin{equation}
\label{dis} \overline{(\Delta N)^{2}} \equiv \overline{(N^{2})}-
(\overline{N})^{2}=\kappa T \Bigl(\frac{\partial
\overline{N}}{\partial \mu}\Bigr)_{T,V},
\end{equation}
where $N$ is the total number of particles of the gas. The
expression (\ref{dis}) is directly related to the isothermal
compressibility $\kappa_{T}$ through \cite{yukalov,Phatria}
\begin{equation}
\label{IK}
 \kappa_{T}=\frac{\overline{(\Delta N)^{2}}}{\rho N \kappa T},
\end{equation}
where $\rho$ is a mean particle density. Using expression
(\ref{NPINT1}), we are able calculated the fluctuations associated
with our modified bosonic gas in the thermodynamic limit, and
analyze, how the deformation parameter $\alpha$ could affect the
stability of the system.
Assuming that $\overline{(\Delta N_{0})^{2}}=0$ and
$\overline{(\Delta N)^{2}}=\overline{(\Delta N_{e})^{2}}$ being
$N_{e}$ the number of particles in the excited states
\cite{yukalov3}, allows us to calculate from (\ref{NPINT1}),
together with (\ref{dis}), the fluctuations associated with our
modified bosonic gas
\begin{equation}
 \label{dis1}
\overline{(\Delta N_{e})^{2}}=\frac{(\kappa T)^{\gamma}}{V_{char}}
\Bigg[\Gamma(3/2)\frac{2 \pi (2m)^{3/2}}{(2 \pi
\hbar)^3}g_{\gamma-1}(z) -\alpha \frac{8 \pi m^{2}}{(2 \pi
\hbar)^{3}}(\kappa T)^{-1/2}g_{\gamma-3/2}(z)\Bigg].
\end{equation}
The relation between the critical temperature and the characteristic
volume (\ref{volumen}), allows us to write (\ref{dis1}) above the
critical temperature ($T>T_{c}$) as follows
\begin{equation}
\label{dis2} \overline{(\Delta
N_{e})^{2}}=\Bigg[\frac{g_{\gamma-1}(z)}{\zeta(\gamma)}
-\alpha\frac{g_{\gamma-3/2}(z)}{\zeta(\gamma)}\Bigg(\frac{8m}{\pi
\kappa
T}\Bigg)^{1/2}\Bigg]\Bigg(\frac{T}{T_{0}}\Bigg)^{\gamma}N,\hspace{1cm}T>T_{c}.
\end{equation}
For $T>T_{c}$, $z<1$ \cite{Phatria}. The fluctuations increase if
$\alpha<0$, conversely, a positive $\alpha$ decreases the
fluctuations. Nevertheless, the corrections caused by the
deformation parameter leaves the fluctuations \emph{normal},
independent of the sign of $\alpha$; in this case, the isothermal
compressibility is always finite and positive for $\gamma>1$,
according to (\ref{IK}). Hence, the system is always stable for
temperatures above $T_{c}$, as is expected.
A more interesting situation is the analysis of the fluctuations for
temperatures $T<T{c}$, in this case the dispersion (\ref{dis2})
becomes
\begin{equation}
\label{dis3} \overline{(\Delta
N_{e})^{2}}=\Bigg[\frac{\zeta(\gamma-1)}{\zeta(\gamma)}
-\alpha\frac{\zeta(\gamma-3/2)}{\zeta(\gamma)}\Bigg(\frac{8m}{\pi
\kappa
T}\Bigg)^{1/2}\Bigg]\Bigg(\frac{T}{T_{0}}\Bigg)^{\gamma}N,\hspace{1cm}T<T_{c}.
\end{equation}
We notice that there is a critical value for the shape parameter
$\gamma$ which suggest that the fluctuations go from being
\emph{normal} to be \emph{anomalous}, this value is $\gamma=5/2$. In
this case, the term proportional to $\alpha$ in (\ref{dis3}) becomes
divergent due to the fact that $\zeta(1)\rightarrow \infty$, hence,
the isothermal compressibility, apparently, becomes divergent for
temperatures $T<T_{c}$. For values, $\gamma \leq 5/2$, the
fluctuations caused by the deformation parameter $\alpha$ becomes
\emph{anomalous}, and the isothermal compressibility becomes,
apparently, divergent according to (\ref{IK}), which means that the
system could be unstable beyond the critical temperature. However,
this happens only in the thermodynamic limit and assuming that $\mu=0$
at the critical temperature $T_{c}$. In real systems the
thermodynamic limit is never reached, and in fact, the isothermal
compressibility is not really divergent. The divergence in the
isothermal compressibility (or this apparently
\emph{anomalous} behavior) can be eliminated by using the lowest
energy associated to the system at the critical temperature
$T_{c}$ \cite{yukalov1}. In other words, $\mu=\epsilon_{min}$ at
$T_{c}$. For instance, we are able to express the term proportional to $\alpha$ in (\ref{IK}) as $g_{\gamma-3/2}(e^{\beta_{c}(\epsilon_{min}-m \alpha^{2}/2)})/\zeta(\gamma)$ and safely remove the divergences, by using the properties of the Bose--Einstein functions when its argument tends to zero \cite{Phatria}. Nevertheless, the previous exposition, suggest that the
associated isothermal compressibility could be very large (o very
small, when $\xi_{1}<0$), for values of the shape parameter $\gamma
\leq 5/2$, as a consequence of the deformation in the dispersion
relation, for temperatures $T<T_{c}$. In more realistic systems, the finite size corrections (or
finite number of particles) and  interatomic interactions must be
taken into account, and could be interesting to analyze the behavior
of the corresponding $\kappa_{T}$. Finally, in this range of values for the
shape parameter $\gamma$ we have, for instance, potentials of the
type $V(\vec{r})\sim x^{3}+y^{3}+z^{3}$ in the case of a Cartesian
traps, $V(\vec{r})\sim \rho^{3}+z^{3}$ in the case of cylindrical
traps, and $V(\vec{r})\sim r^{3}$ in the case of spherical traps,
also for the potentials, $V(\vec{r})\sim \rho^{4}+z^{2}$,
$V(\vec{r})\sim r^{6}$ or any other combination such that $\gamma
\leq 5/2$.

\section{Conclusions}
 By using the formalism of the semiclassical approximation, we have analyzed the Bose--Einstein condensation for modified bosonic gas trapped in a 3--D power law potential. We have proved that the critical temperature must be corrected as a consequence of the deformation in the dispersion relation. The shift in the critical temperature caused by such deformation, expressed in terms of the number of particles and the trap parameters can be used to provided bounds for the deformation parameter $\xi_{1}$, wich in the case of typical laboratory conditions allows to bound such parameter up to $|\xi_{1}|\lesssim10^{4}$, for a harmonic oscillator potential. Moreover, we have stressed that the shift in the condensation temperature could be enhanced by taking systems containing a finite number of particles and/or a massive bosons and/or lower frequencies, trapped in a quite general potential, and that these systems can be used to improve the bounds associated to the deformation parameter. Additionally, the finite size effects and the interaction among the particles, affect the condensation temperature, and  must be taken into account. The analysis of the condensation temperature of a modified bosonic gas in systems containing a finite number of particles (finite size corrections), in the interacting and non--interacting cases is analyzed in \cite{CastellanosClaus}. Finally, the present work, opens the possibility to study the corrections caused by an anomalous dispersion relation over the main properties associated to the condensate, within the Bogoliubov formalism (see for example \cite{Pethick,AAA}) for instance, the speed of sound, the pressure and energy associated to the ground state, the healing length, among others and deserves a detailled investigation \cite{eli}, in order analyze the possibilities to improve the results given in this work.
\begin{acknowledgments}
This research was supported by DAAD (Deutscher Akademischer
Austauschdienst) under grant $A/09/77687$.
\end{acknowledgments}

\end{document}